%% file: ms.tex
\pdfoutput=1
\PassOptionsToPackage{dvipsnames}{xcolor}

\documentclass[sigplan]{acmart}

\usepackage{algorithm}
\usepackage[noend]{algpseudocode}
\usepackage{cleveref}
\usepackage{listings}

\usepackage{dcolumn}
\usepackage{multirow}
\usepackage{numprint}
\usepackage{xfp}

\usepackage{xspace}

\usepackage[htt]{hyphenat}

\makeatletter
\DeclareRobustCommand\ttfamily{%
  \not@math@alphabet\ttfamily\mathtt\fontfamily\ttdefault%
  \small%
  \selectfont}
\makeatother

\crefname{lstlisting}{Listing}{Listings}
\crefname{algorithm}{Algorithm}{Algorithms}

\title{An Ownership Policy and Deadlock Detector for Promises}

\author{Caleb Voss}
\email{cvoss@gatech.edu}
\affiliation{%
  \institution{Georgia Institute of Technology}
}

\author{Vivek Sarkar}
\email{vsarkar@gatech.edu}
\affiliation{%
  \institution{Georgia Institute of Technology}
}

\begin{document}

\setcopyright{none}
\renewcommand\footnotetextcopyrightpermission[1]{}
\pagestyle{plain}

\input{listings}

\input{algorithms}
\input{tables}

\begin{abstract}
  Task-parallel programs often enjoy deadlock freedom under certain
  restrictions, such as the use of structured join operations, as in
  Cilk and X10, or the use of asynchronous task futures together with
  deadlock-avoiding policies such as Known Joins or Transitive Joins.
  However, the \emph{promise}, a popular synchronization primitive for
  parallel tasks, does not enjoy deadlock-freedom guarantees. Promises
  can exhibit deadlock-like bugs; however, the concept of a deadlock
  is not currently well-defined for promises.

  To address these challenges, we propose an ownership semantics in
  which each promise is associated to the task which currently intends
  to fulfill it.
  Ownership immediately enables the identification of bugs in which a
  task fails to fulfill a promise for which it is responsible.
  Ownership further enables the discussion of deadlock cycles among
  tasks and promises and allows us to introduce a robust definition of
  deadlock-like bugs for promises.

  Cycle detection in this context is non-trivial because it is
  concurrent with changes in promise ownership.
  We provide a lock-free algorithm for precise runtime deadlock
  detection.
  We show how to obtain the memory consistency criteria required for
  the correctness of our algorithm under TSO and the Java and C++
  memory models.
  An evaluation compares the execution time and memory usage overheads
  of our detection algorithm on benchmark programs relative to an
  unverified baseline. Our detector exhibits a \geomeanTimeFancy
  geometric mean time overhead and a \geomeanMemFancy geometric mean
  memory overhead, which are smaller overheads than in past approaches
  to deadlock cycle detection.
\end{abstract}

\maketitle

\section{Introduction}

The task-parallel programming model is based on the principle that
structured parallelism (using high-level abstractions such as
spawn-sync~\cite{Cilk,OpenMP}, async-finish~\cite{X10,HJlib,hclib},
futures~\cite{JCP,Cpp17}, barriers~\cite{OpenMP}, and
phasers~\cite{HJ,Phaser}) is a superior style to unstructured
parallelism (using explicit low-level constructs like threads and
locks).
Structured programming communicates programmer intent in an upfront
and visible way, providing an accessible framework for reasoning about
complex code by isolating and modularizing
concerns.
However, the \emph{promise} construct, found in mainstream languages
including C++ and Java, introduces an undesirable lack of structure
into task-parallel programming. A promise generalizes a future in that
it need not be bound to the return value of a specific task.
Instead, any task may elect to supply the value, and the code may not
clearly communicate which task is intended to do so.

Promises provide point-to-point synchronization wherein one or more
tasks can await the arrival of a payload, to be produced by another
task.
Although the promise provides a safe abstraction for sharing data
across tasks, there is no safety in the kinds of inter-task blocking
dependencies that can be created using promises.
The inherent lack of structure in promises not only leads to
deadlock-like bugs in which tasks block indefinitely due to a cyclic
dependence, but such bugs are not well-defined and are undetectable in
the general case due to the lack of information about which task is
supposed to fulfill which promise.

\lstDeadlock

A deadlock-like cycle may only be detected once all tasks have
terminated or blocked. For example, the Go language runtime reports a
deadlock if no task is eligible to run~\cite{Go}. However, if even one
task remains active, this technique cannot raise an alarm.
An example of such a program is in \cref{lst:deadlock}; the root task
and $t_2$ are in a deadlock that may be hidden if $t_1$ is a
long-running task, such as a web server.
An alternative detection approach is to impose timeouts on waits,
which is only a heuristic solution that may raise an alarm when there
is no cycle.
In both of these existing approaches, the detection mechanism may find
the deadlock some time \emph{after} the cycle has been created.
It is instead more desirable to detect a cycle immediately when it
forms.

\subsection{Promise Terminology}

There is inconsistency across programming languages about what to call
a promise and sometimes about what functionality ``promise'' refers
to.
The synchronization primitive we intend to discuss is called by many
names, including promise~\cite{Cpp17}, handled
future~\cite{LambdaFut}, completable
future~\cite{JavaCompletableFuture}, and one-shot
channel~\cite{RustOneshot}.
For us, a promise is a wrapper for a data payload that is initially
absent; each \emph{get} of the payload blocks until the first
and only \emph{set} of the payload is performed.
Setting the payload may also be referred to as completing, fulfilling,
or resolving the promise.

Some languages, such as C++, divide the promise construct into a pair
of objects; in this case, ``promise'' refers only to the half with a
setter method, while ``future'' refers to the half with a getter
method.
In Java, the \textsf{CompletableFuture} class is a promise, as it
implements the \textsf{Future} interface and additionally provides a
setter method.

Habanero-Java introduced the data-driven future~\cite{DDF}, which is a
promise with limitations on when gets may occur. When a new task is
spawned, the task must declare up front which promises it intends to
consume. The task does not become eligible to run until all such
promises are fulfilled.

In JavaScript, the code responsible for resolving a promise must be
specified during construction of the
promise~\cite{JavaScriptPromise}. This is a limitation that makes
deadlock cycles impossible, although the responsible code may omit to
resolve the promise altogether, leading to unexecuted callbacks.

Promises may provide a synchronous or an asynchronous API. The Java
concurrency library provides both, for
example~\cite{JavaCompletableFuture}.
The synchronous API consists of the \emph{get} and \emph{set} methods.
The asynchronous API associates each of the synchronous operations to a
new task.
A call to \emph{supplyAsync} binds the eventual return value of a new
task to the promise.
The \emph{then} operation schedules a new task to operate on the
promise's value once it becomes available.
The asynchronous API can be implemented using the synchronous API.
Conversely, the synchronous API can be implemented using continuations
and an asynchronous event-driven scheduler \cite{Imam14}.
We focus on the synchronous API in this work.

\subsection{Two Bug Classes}

We identify two kinds of synchronization bug in which the improper use
of promises causes one or more tasks to block indefinitely:
\begin{enumerate}
\item the \emph{deadlock cycle}, in which tasks are mutually blocked
  on promises that would be set only after these tasks unblock, and
\item the \emph{omitted set}, in which a task is blocked on a promise
  that no task intends to set.
\end{enumerate}
However, neither of these bugs manifests in an automatically
recognizable way at runtime unless every task in the program is
blocked.
In fact, the definitions of these bugs describe conditions which
cannot generally be detected.
What does it mean for no task to \emph{intend} to set a promise?
What does it mean that a task \emph{would} set a promise once the task
unblocks?
In a traditional deadlock, say one involving actual locks, the cycle
is explicit: Task 1 holds lock $A$ and blocks while acquiring lock
$B$, because task 2 is holding lock $B$ and concurrently blocked
during its acquisition of lock $A$.
Intention to release a lock (thereby unblocking any waiters) is
detectable by the fact that a task holds the lock.
But we currently have no concept of a task ``holding'' a promise and
no way to tell that a task intends to set it.

\newcommand\defkw[1]{
  \expandafter\newcommand\csname kw#1\endcsname{%
    \textsf{\textbf{#1}}\ifmmode\ \else\xspace\fi%
  }
}
\defkw{new}
\defkw{set}
\defkw{get}
\defkw{async}

\newcommand\deffld[1]{
  \expandafter\newcommand\csname fld#1\endcsname{%
    \textsf{#1}\xspace%
  }
}
\deffld{owner}
\deffld{owned}
\deffld{waitingOn}

\newcommand\Null{\mathit{null}}

\subsection{Need for Ownership Semantics}

Consider the small deadlock in \cref{lst:deadlock}.
Two promises, $p,q$, are created.
Task $t_2$ waits for $p$ prior to setting $q$, whereas the root task
waits for $q$ prior to setting $p$.
Clearly a deadlock cycle arises? Not so fast. To accurately call this
pattern a deadlock cycle requires knowing that task $t_1$ will not
ever set $p$ or $q$. Such a fact about what \emph{will} not happen is
generally not determinable from the present state without an offline
program analysis.
For this reason, a deadlock cycle among promises evades runtime
detection unless the cycle involves every currently executing task.

\lstOmittedSet

Now consider the bug in \cref{lst:omittedset}.
Two promises, $r,s$, are created. According to the comments, task
$t_3$ is responsible for setting both, and it subsequently delegates
the responsibility for $s$ to $t_4$.
However, $t_4$ fails to perform its intended behavior, terminating
without setting $s$.
The root task then blocks on $s$ forever.
If a bug has occurred, we would like to raise an alarm at runtime when
and where it occurs.
Where is this bug? Should the root task not have blocked on $s$?
Should $t_4$ have set $s$? Should $t_3$ have set $s$?
The blame cannot be attributed, and the bug may, in fact, be in any
one of the tasks involved.
Furthermore, \emph{when} does this bug occur?
The \emph{symptom} of the bug manifests in the indefinite blocking of
the root task, potentially \emph{after} $t_4$ terminates successfully.
If some other task may yet set $s$, then this bug is not yet confirmed
to have occurred.
Omitted sets evade runtime detection and, even once discovered, evade
proper blame assignment.

We propose to augment the task creation syntax (\kwasync in our
examples) to carry information about promise ownership and
responsibility within the code itself, not in the comments.
In doing so, omitted sets become detectable at runtime with blame
appropriately assigned.
Moreover, programmer intent is necessarily communicated in the code.
Finally, in knowing which task is expected to set each promise, it
becomes possible to properly discuss deadlock cycles among promises.

\subsection{Omitted Set in the Wild}

\lstAmazon

An example of an omitted set bug was exhibited by the Amazon Web
Services SDK for Java (v2) when a certain checksum validation
failed~\cite{AWSBugReport}.
An abbreviated version of the code is given in \cref{lst:amazon};
line~\ref{ln:amazon:fix} was absent prior to the bug fix.
The control flow ensures that either exception handling code or
non-exceptional code was executed, not both
(line~\ref{ln:amazon:bug})~\cite{AWSBugIntroduced}.
However, only the non-exceptional code would set the value of a
\texttt{CompletableFuture} (Java's promise) to indicate the work was
completed (line~\ref{ln:amazon:complete}), whereas the
\textsf{onError} method would take no action.
If checksum validation failed after a file download, any consumer
tasks waiting for the download to complete would block indefinitely.
A month later, the omitted set bug was identified and
corrected by adding line~\ref{ln:amazon:fix}~\cite{AWSBugFixed}.

When this bug arises at runtime, the symptom (the blocked consumer) is
far from the cause (the omitted set), and the bug is not readily
diagnosable.
If the runtime could track which tasks are responsible for which
promises, then this bug could be detected and reported as an exception
as soon as the responsible task terminates.
Using our approach, the bug would be detected when the task running
the \texttt{onComplete} callback finishes, and the alarm would name
the offending task and the unfulfilled promise.

\subsection{Contributions}

In this work, we propose the addition of \emph{ownership semantics} for
promises which enables a task's intention to set a promise to be
reflected in the runtime state.
In so doing,
\begin{enumerate}
\item we enable a precise definition of a deadlocked cycle of promises
  in terms of runtime state;
\item we define a second kind of blocking bug, the \emph{omitted
    set}, which does not involve a cycle;
\item we require important programmer intent to be encoded explicitly
  and to respect a runtime-verifiable policy, thereby enabling
  structured programming for promises.
\end{enumerate}
In addition to these theoretical contributions,
\begin{enumerate}
\item we introduce a new lock-free algorithm for detecting our
  now-identifiable deadlock-cycle and omitted-set bugs \emph{when they
    occur};
\item we identify properties critical for establishing the correctness
  of the algorithm under weak memory consistency and show how to
  ensure these properties hold under the TSO, Java, and C++ memory
  models;
\item we prove that our algorithm precisely detects every deadlock
  without false alarms;
\item we experimentally show that a Java implementation has low
  execution time and memory usage overheads on nine benchmarks
  relative to the original, unverified baseline (geometric mean
  overheads of \geomeanTime and \geomeanMem, respectively).
\end{enumerate}

\input{ownership}

\input{detector}
\input{weak}
\input{correctness}

\input{evaluation}
\input{related}

\section{Conclusion}

We have introduced an ownership mechanism for promises, whereby each
task is responsible for ensuring that all of its owned promises are
fulfilled.
This mechanism makes it possible to identify a bug, called the omitted
set, at runtime when the bug actually occurs and to report which task
is to blame for the error.
The ownership mechanism also makes it meaningful, for the first time,
to formally define, discuss, and detect deadlock cycles among tasks
synchronizing with promises. Such a bug is now detectable as soon as
the cycle forms.

In our approach, any code that spawns a new asynchronous task must
name the promises which are to be transferred to the new task.
The programmer must already be aware of this critical information in
order to even informally reason about omitted set and deadlock
bugs. We now ask that it be explicitly notated in the code.

We provided an algorithm to check for compliance with the ownership
policy at runtime, thereby detecting omitted sets, as well as an
algorithm for detecting deadlock cycles using ownership information.
Both types of bug are detected when they occur, not after-the-fact.
Our deadlock detector is provably precise and correct under a weak
memory model and we described how to obtain this correct behavior
under the TSO, Java, and C++ memory models. Every alarm corresponds to
a true deadlock and every deadlock results in an alarm.
Experimental evaluation demonstrates that our lock-free approach to
deadlock detection exhibits low execution time and memory
overheads relative to an uninstrumented baseline.

\balance

\begin{acks}
  This work is supported by the \grantsponsor{NSF}{National Science
    Foundation}{https://www.nsf.gov} under Collaborative Grant
  No.~\grantnum{NSF}{1822919} and Graduate Research Fellowship Grant
  No.~\grantnum{NSF}{1650044}.
\end{acks}

\bibliography{ref}

\end{document}

%% file: listings.tex
\lstset{
  float,
  belowskip=-\baselineskip,
  escapechar=\#,
  numbers=left,
  xleftmargin=15pt,
  numbersep=5pt,
  language=Java,
  basicstyle=\footnotesize\sffamily,
  commentstyle=\color{MidnightBlue},
  keywordstyle=[1]\bfseries,
  morekeywords=[1]{async},
  keywordstyle=[2]\itshape,
  morekeywords=[2]{},
  keywordstyle=[3]\bfseries\color{RedOrange},
  morekeywords=[3]{put,set,complete,completeExceptionally},
  keywordstyle=[4]\bfseries\color{LimeGreen},
  morekeywords=[4]{get,join},
  keywordstyle=[5]\bfseries\color{Magenta},
  morekeywords=[5]{guard}
}

\makeatletter
\let\lst@floatdefault\lst@float
\makeatother

\def\lstDeadlock{
  \lstinputlisting[
  caption={A deadlock?},
  label={lst:deadlock}]
  {listings/deadlock.java}
}

\def\lstOmittedSet{
  \lstinputlisting[
  caption={An omitted set?},
  label={lst:omittedset}]
  {listings/omittedset.java}
}

\def\lstChannel{
  \lstinputlisting[
  caption={Object-oriented approach to promise movement.},
  label={lst:channel}]
  {listings/channel.java}
}

\def\lstAmazon{
  \lstinputlisting[
  caption={An omitted set in Amazon AWS SDK (v2)~\cite{AWSBugReport}.
    Code abbreviated and inlined for clarity.},
  label={lst:amazon}]
  {listings/amazon.java}
}

%% file: ownership.tex
\section{Ownership Policy}

\def\Lp{\ensuremath{\mathcal{L}_p}\xspace}
\def\Po{\ensuremath{\mathcal{P}_o}\xspace}

In promise-based synchronization, a task does not directly await
another task; it awaits a promise, thereby \emph{indirectly} waiting
on whichever task fulfills that promise.
It is a runtime error to fulfill a promise twice, so there ought to be
one and only one fulfilling task.
However, the relationship between a promise and the task which
\emph{will} fulfill it is not explicit and inhibits the identification
of deadlocks.
To make this relationship explicit and meaningful, we say that each
promise is \emph{owned} by exactly one task at any given time.
The owner is responsible for fulfilling the promise eventually, or
else handing ownership off to another task.
Ownership hand-offs may only occur at the time of spawning a new
task.
We augment the \kwasync keyword, used to spawn tasks, with a list of
promises currently owned by the parent task that should be transferred
to the new child.

\subsection{Language Extension}

We define an abstract language, showing only its synchronization
instructions and leaving its sequential control flow and other
instructions unspecified.
For simplicity, we have abstracted away the payload values of promises
and refer to individual promises by globally unique identifiers.
\begin{definition}
  The \Lp language consists of task-parallel programs, $P$, whose
  synchronization instructions have the syntax
  \begin{align*}
    \kwnew p ~|~ \kwset p ~|~ \kwget p
    ~|~ \kwasync(p_1,\ldots,p_n)\ \{ P \}
  \end{align*}
  where $n$ may be $0$.
\end{definition}
The instruction $\kwnew p$ represents the point of allocation for the
promise $p$, and we assume well-formed programs do not allocate a
given $p$ twice or operate on $p$ prior to its allocation.
Each invocation of $\kwget p$ blocks the current task until after
$\kwset p$ has been invoked for the first (and only) time.

The \kwasync block creates a new task to execute a sub-program $P$;
the block is annotated with a list of promises, which should be moved
from the parent task to the new task.
In many task-parallel languages, \kwasync automatically creates a
future which can be used to retrieve the new task's return value. We
can readily reproduce this behavior using promises in the pattern
$\kwnew p; \kwasync (p, \ldots)~\{ \ldots; \kwset p \}$.

\begin{definition}
  The \emph{ownership policy}, \Po, maintains state during the
  execution of an \Lp program in the form of a map
  $\fldowner : \mathit{Promise} \to \mathit{Task} \cup \{ \Null \}$
  according to these rules:
  \begin{enumerate}
  \item When task $t$ executes $\kwnew p$, set $\fldowner(p) := t$.
    \label{rule:new}

  \item When task $t$ spawns task $t'$ as
    $\kwasync(p_1,\ldots,p_n)\ \{ P \}$, prior to $t'$ becoming
    eligible to run, ensure $\fldowner(p_i) = t$ and update
    $\fldowner(p_i) := t'$ for each $p_i$.
    \label{rule:async}

  \item When task $t$ terminates, ensure the set of promises
    $\fldowner^{-1}(t)$ is empty.
    \label{rule:exit}

  \item When task $t$ executes $\kwset p$, ensure that
    $\fldowner(p) = t$ and set $\fldowner(p) := \Null$.
    \label{rule:set}
  \end{enumerate}
\end{definition}

These four rules together ensure that there is at least one \kwset for
each promise, with omitted sets being detected by
rule~\ref{rule:exit}. Rule~\ref{rule:set} guarantees there is at most
one \kwset.

Our proposed modification to the program given in \cref{lst:deadlock}
is to annotate the \kwasync in line~\ref{ln:deadlock:t2} as
$\kwasync(q)$, indicating that $t_2$ takes on the responsibility to
set $q$.
It is now possible to trace the cycle when it occurs: the root task
awaits $q$, owned by $t_2$, awaiting $p$, owned by the root task. It
is clear that $t_1$, whose \kwasync is not given any parameters, is
not involved as it can set neither $p$ nor $q$ (rule~\ref{rule:set}).

The proposed modification to the program given in
\cref{lst:omittedset} is to write $\kwasync(r,s)$ in
line~\ref{ln:omittedset:t3} and $\kwasync(s)$ in
line~\ref{ln:omittedset:t4}. That is, the information already present
in the comments is incorporated into the code itself.
The moment $t_4$ terminates, the runtime can observe that $t_4$ still
holds an outstanding obligation to set $s$. We treat this as an error
immediately (rule~\ref{rule:exit}), irrespective of whether any task
is awaiting $s$.

\subsection{Algorithm for Ownership Tracking}

\algOwners

\Cref{alg:owners} implements the \Po policy by providing code to be
run during \kwnew, \kwasync, and \kwset operations.
Each promise has an $\fldowner$ field to store the task that is
currently its owner, and each task has an associated $\fldowned$
list that maintains the inverse map, $\fldowner^{-1}$.
The functions $\mathit{currentTask}$ and $\mathit{getCurrentTask}$
interact with thread-local storage.

In compliance with \Po rule~\ref{rule:new}, the \textsc{New}
procedure creates a promise owned by the currently running task
(line~\ref{ln:owners:new:A}) and adds this promise to that task's
owned list (line~\ref{ln:owners:new:B}).

$\textsc{Async}(P, f)$ schedules $f$ to be called asynchronously as a
new task and moves the promises listed in $P$ into this task.
These promises are first confirmed to belong to the parent task
(line~\ref{ln:owners:async:A}), then moved into the child task
(lines~\ref{ln:owners:async:A}--\ref{ln:owners:async:C}), in
accordance with rule~\ref{rule:async}.
(Line~\ref{ln:owners:async:B} is in preparation for
\cref{alg:detector}, presented in \cref{sec:detector}.)
Once the child task terminates, rule~\ref{rule:exit} requires that the
task not own any remaining promises (line\ref{ln:owners:async:E}).
The \textsc{Init} procedure shows how to set up a root task to execute
the main function.

Finally, $\textsc{Set}(p,v)$ achieves rule~\ref{rule:set}, checking
that the current task owns $p$ and marking $p$ as fulfilled by
assigning it no owner
(lines~\ref{ln:owners:set:A}--\ref{ln:owners:set:C}).
The procedure then invokes the underlying mechanism for actually
setting the promise value to $v$ (line~\ref{ln:owners:set:D}).

As an example of how \Cref{alg:owners} enforces compliance with \Po,
refer again to \cref{lst:omittedset}. When promise $s$ is first
created, it belongs to the root task (\cref{alg:owners}
\cref{ln:owners:new:B}). If the \kwasync that creates $t_4$ is
annotated with $s$, then \cref{alg:owners} \cref{ln:owners:async:C}
changes the owner of $s$ to $t_4$. Since $t_4$ does not set $s$, upon
termination of $t_4$, an assertion fails in \cref{alg:owners}
\cref{ln:owners:async:E}.
The offending task, $t_4$, and the outstanding promise, $s$, are
directly identifiable and can be reported in the alarm.

%% file: detector.tex
\section{Deadlock Detection Algorithm}
\label{sec:detector}

Now that we have established the relationship between promises and
tasks, it is possible to describe what a deadlock is.
A deadlock is a cycle of $n$ tasks, $t_i$, and $n$ promises, $p_i$,
such that $t_i$ awaits $p_i$ while $p_i$ is owned by $t_{i+1}$
($\mathrm{mod}~n$).
The information required to identify such a deadlock is, for the first
time, made available explicitly at runtime through the use of the \Po
policy.
We can now develop a runtime detection mechanism to identify deadlocks
based on this information and raise an alarm as soon as one is
created.

\subsection{Approach}

Even assuming sequential consistency, the algorithm for finding such a
cycle is non-trivial. Conceptually, whenever a $\kwget p$ is executed
by $t$, $t$ must alternately traverse owned-by and waits-for edges to
see if the path of dependences returns to $t$. If another task, $t'$,
is encountered which is not currently awaiting a promise, this proves
that progress is still being made and there is no deadlock (yet). In
this case, $t$ passes verification and commits to blocking on
$p$. Should this path of dependences grow due to a subsequent
$\kwget p'$ by $t'$, then the same algorithm runs again in task $t'$
to verify that the new waits-for edge does not create a deadlock.

Crucially, during verification $t$ must establish a waits-for edge to
mark that it is awaiting $p$ \emph{prior} to traversing the dependence
path. That is, a waits-for edge is created before it is determined
that $t$ will be allowed to await $p$. A two-task cycle shows what
would go wrong if this procedure is not followed. If $t$ begins to
verify its wait of $p$ (say, owned by $t'$) without marking that $t$
is awaiting $p$, and concurrently $t'$ begins to verify its wait of
$p'$ (owned by $t$) without marking that $t'$ is awaiting $p'$, then
each task may find that the other is apparently not awaiting any
promises at this time, and both commit to blocking, creating an
undetected deadlock. However, by ensuring that each task marks itself
as awaiting a promise prior to verifying whether that wait is safe, we
guarantee the last task to arrive in the formation of a deadlock cycle
will be able to detect this cycle.

A second consideration is how this approach handles concurrent
transfer of promise ownership or concurrent fulfillment of
promises. Suppose that while the cycle detection algorithm is
traversing a dependence path, an earlier promise in the path is
transferred to a new owner or is fulfilled, thereby invalidating the
remainder of the traversed path. Failure to handle this correctly
could result in an alarm when there is no deadlock. The first
observation we make is that this scenario cannot arise for any but the
most recent promise encountered on the path. If $p_0$ is owned by
$t_1$, awaiting $p_1$, owned by $t_2$, then it is impossible for $p_0$
to move into a new task or to become fulfilled, since its current
owner, $t_1$, is blocked (or about to block, pending successful
verification). The concern is only that $t_2$ has not yet blocked and
may transfer or fulfill $p_1$. The natural solution is that when
traversing the dependence path, upon reaching each promise in the path
we must go back and double-check that the \emph{preceding} promise
still belongs to the task it belonged to in the previous iteration and
is still unfulfilled. If this check fails, then the present
verification passes because progress is still being made.

\subsection{Detection Algorithm}

\algDetector

The deadlock detector occupies the implementation of the \kwget
instruction, given in \cref{alg:detector}.
This detector can thereby raise an alarm in a task as soon as the task
attempts a deadlock-forming await of a promise.
At the time of raising an alarm, the available diagnostic information
that can be reported includes the task, the awaited promise, as well
as every other task and promise in the cycle, if desired.

For a preliminary understanding of the procedure's logic, we assume
sequential consistency in this section.
Upon entering \textsc{Get}, the currently executing task records the
promise that it will be waiting on (line~\ref{ln:detector:enter}).
This $\fldwaitingOn$ field was initialized to $\Null$ in
\cref{alg:owners} line~\ref{ln:owners:async:B}, and is always reset to
$\Null$ upon exiting \textsc{Get} (\cref{alg:detector}
line~\ref{ln:detector:final}), either normally
(line~\ref{ln:detector:return}) or abnormally
(line~\ref{ln:detector:fail}). Doing so makes the algorithm robust to
programs with more than one deadlock.

The loop in the detection algorithm traverses the chain of alternating
$\fldowner$ and $\fldwaitingOn$ fields.
If task $t$ is waiting on promise $p$, which is owned by a task $t'$,
then $t$ is effectively waiting on whatever $t'$ awaits.
In traversing this chain, if $t$ finds that it is transitively waiting
on itself, then we have identified a deadlock
(lines~\ref{ln:detector:loop},~\ref{ln:detector:fail}).
If the algorithm reaches the end of this chain without finding $t$
again, as indicated by finding a $\Null$ value in
line~\ref{ln:detector:breakt} ($p_i$ is already fulfilled) or in
line~\ref{ln:detector:breakp} ($t_{i+1}$ is not awaiting a promise),
then it is safe to commit to a blocking wait on the desired promise
(line~\ref{ln:detector:return}).
Recall that $p_i.\fldowner$ is $\Null$ after $p_i$ has been fulfilled,
and $t_{i+1}.\fldwaitingOn$ is $\Null$ when $t_{i+1}$ is not currently
executing \textsc{Get}.

In order to guarantee that an apparent cycle always corresponds to a
real deadlock, even under concurrent updates to promises, we rely on
line~\ref{ln:detector:changed} to establish that task $t_{i+1}$ was
waiting on promise $p_{i+1}$ \emph{while} $t_{i+1}$ was still the
owner of promise $p_i$.
This is achieved by reading the $\fldowner$ field both before
(line~\ref{ln:detector:owner1},~\ref{ln:detector:owner2}) and after
(line~\ref{ln:detector:changed}) reading the $\fldwaitingOn$ field
(line~\ref{ln:detector:waitingOn}).
If the task observes the owner of $p_i$ to have changed, it turns out
that it is safe to abandon the deadlock check and commit to the
blocking wait.

In \crefrange{sec:def}{sec:correct}, we will move to a weaker memory
model. There are two crucial points to remember. We must preserve the
ability to reason temporally over the edges in the dependence path,
and we must guarantee that at least one task entering a deadlock can
observe the existence of the whole deadlock cycle.

%% file: weak.tex
\section{Weakly Consistent Definition of Deadlock}
\label{sec:def}

With a few tweaks, we can obtain a correctness guarantee for our
deadlock detector under a weak memory model, which implies the same
guarantee under any stronger model, including sequential consistency.
First, we must define this weak memory model and give a definition of
deadlock that is compatible with it.

In practice, we do not want to assume that maps such as the
$\fldowner$ field have a single, globally consistent state that is
observed by all tasks. Machines and languages often have weaker
consistency guarantees, and there are performance costs for requesting
stronger consistency due to the synchronization required.
Instead, we will assume a weak memory model and use unsynchronized
accesses whenever possible.

We now define this weak memory model, which we will use to establish
the correctness of our deadlock detection algorithm under models at
least as strong as this one.
\begin{definition}
  The \emph{happens-before} (h.b.) order is a partial order over the
  instructions in a program execution that subsumes the intra-task
  program order and, upon spawning each new task, the ordering of
  \cref{alg:owners} line~\ref{ln:owners:async:D} (the start of the new
  task) after \cref{alg:owners} line~\ref{ln:owners:async:C} (the last
  action of the parent task before spawning).
  The reverse of happens-before is \emph{happens-after}.
\end{definition}
\begin{definition}
  With respect to a given memory location, a read may only
  \emph{observe} a (not necessarily unique) last write which
  happens-before it or any write with which the read is not
  h.b.~ordered.
  Two writes or a write and read of the same location which are not
  h.b.~ordered are \emph{racing}.
\end{definition}
A typical language has a more refined happens-before ordering and
definition of observable writes, especially relating to reads-from
edges on promises; however, we will not need to appeal to such edges
in our formalism.
\begin{definition}
  A program in \Lp is \emph{well-formed} if, in every execution, for
  each promise, $p$, there is at most one $\kwnew~p$ instruction, and
  each \kwset, \kwget, or \kwasync instruction referring to $p$
  happens-after such a $\kwnew~p$.
  \label{def:wf}
\end{definition}

We note that although the owners of different promises may be updated
concurrently, it is not possible in \cref{alg:owners} for a
write-write race to occur on the same owner field.
\begin{lemma}
  Consider an execution of a well-formed program. If $w_1, w_2$ are
  two writes to $p.\fldowner$ in \cref{alg:owners}, then $w_1$ and
  $w_2$ are not racing.
  Further, if $r$ is a read of $p.\fldowner$ by task $t$, and $r$
  observes the value to be $t$, then $r$ does not race with the write
  it observes.
  \label{lem:writeorder}
\end{lemma}
\begin{proof}
  The two claims can be shown together.
  Line~\ref{ln:owners:new:A} represents the initialization of the
  $\fldowner$ field and so happens-before every other write to it.
  The writes in lines~\ref{ln:owners:async:C}
  and~\ref{ln:owners:set:B} each happen-after a read of the same field
  observes the value to be the currently executing task
  (lines~\ref{ln:owners:async:X},~\ref{ln:owners:set:A}).
  Take this together with the fact that there are only two ways to set
  $p.\fldowner$ to $t$: line~\ref{ln:owners:new:A}, executed by $t$
  itself, or line~\ref{ln:owners:async:C}, executed by the parent of
  $t$ prior to spawning $t$. In either case, writing $t$ to
  $p.\fldowner$ happens-before any read of $p.\fldowner$ by $t$
  itself.
\end{proof}

Since we do not assume a globally consistent state, we have to be
careful in the definition of deadlock cycle. Two tasks need not agree
on the value of $\fldowner(p)$ for a given promise, $p$.
Instead of freely referring to $\fldowner$ as a map
$\mathit{Promise} \to \mathit{Task} \cup \{ \Null \}$, we must
additionally state which task's perspective is being used to observe
the $\fldowner$ map.
\begin{definition}
  A non-empty set of tasks, $T$, is in a \emph{deadlock cycle} if
  for every task $t \in T$,
  \begin{enumerate}
  \item $t$ is executing $\kwget p_t$ for some promise, $p_t$,
  \item there exists a task, $o_{p_t}$, also in $T$ which observes that
    $\fldowner(p_t) = o_{p_t}$,
  \end{enumerate}
  and $T$ is minimal with respect to these constraints.
  The set of promises associated to the deadlock is
  $\{ p_t ~|~ t \in T \}$.
\end{definition}
The subtle point in this definition is that task $o_{p_t}$ necessarily
has the most up-to-date information about the owner of $p_t$, since
$o_{p_t}$ is itself the owner. Per \cref{lem:writeorder}, we know that
all the writes to $p_t.\fldowner$ are ordered and that $o_{p_t}$ is
observing the last such write, since only $o_{p_t}$ is capable of
performing the next write to follow the observed one.

%% file: correctness.tex
\section{Correctness under Weak Consistency}
\label{sec:correct}

\Cref{alg:detector} correctly and precisely detects all deadlocks
under our weak memory consistency model with some additional specific
consistency requirements on certain accesses. We define these
requirements, show how to meet them in each of the TSO, Java, and C++
memory models, and then prove the algorithm raises an alarm exactly
when there is a deadlock.

\subsection{Requirements}

In order to prove correctness, we require the following additional
memory consistency.
\begin{enumerate}
\item There is a total order, $<$, over all instances of the write in
  \cref{alg:detector} line~\ref{ln:detector:enter}, across all memory
  locations. Let $w_1 < w_2$. Any write preceding and including $w_1$
  in h.b.~order is visible to any read following $w_2$ in h.b.~order.
  \label{rule:seq}

\item The consistency of any $\fldowner$ field is expected to
  follow from release-acquire semantics for any $\fldwaitingOn$
  field.
  Specifically, let $w_1$ be an \cref{alg:owners}
  line~\ref{ln:owners:new:A} or line~\ref{ln:owners:async:C} write to
  an $\fldowner$ field, let $w_2$ be an \cref{alg:detector}
  line~\ref{ln:detector:enter} write to a $\fldwaitingOn$ field, let
  $r_2$ be an \cref{alg:detector} line~\ref{ln:detector:waitingOn}
  read, and let $r_1$ be an \cref{alg:detector}
  line~\ref{ln:detector:changed} read.
  Suppose $w_1,r_1$ refer to the same location, as do $w_2,r_2$.
  If $w_1$ happens-before $w_2$, if $w_2$ is visible to $r_2$, and if
  $r_2$ happens-before $r_1$, then $w_1$ is visible to $r_1$.
  \label{rule:acq}

\item The write in \cref{alg:detector} line~\ref{ln:detector:final}
  must not become visible until the fulfillment of $p_0$ is visible
  (\cref{alg:owners} line~\ref{ln:owners:set:B}) or it is determined
  that an exception should be raised (\cref{alg:detector}
  line~\ref{ln:detector:fail}).
  \label{rule:rel}
\end{enumerate}

These three requirements are readily attained in TSO, Java, and C++ as
follows.
\begin{itemize}
\item Under TSO, a memory fence is needed in \cref{alg:detector}
  line~\ref{ln:detector:fence} to achieve requirement~\ref{rule:seq}
  by ordering line~\ref{ln:detector:waitingOn} after
  line~\ref{ln:detector:enter} and sequentializing all instances of
  line~\ref{ln:detector:fence} with each other.
  TSO naturally achieves requirement~\ref{rule:acq} by respecting the
  local store order, as well as requirement~\ref{rule:rel} by not
  allowing the line~\ref{ln:detector:final} write to become visible
  early.
  Note that the loop contains no fences.

\item Under the Java memory model, it suffices to mark the two fields,
  $\fldowner$ and $\fldwaitingOn$, as volatile to satisfy
  all three requirements.
  This eliminates all write-read data races.
  Remember that there are no write-write races (see
  \cref{lem:writeorder}).
  In the absence of any races on these two fields, the Java memory
  model guarantees sequential consistency with respect to these
  fields.

\item In C++ both of the fields must be \texttt{std::atomic} to
  eliminate data races, but this alone is insufficient.
  \Cref{alg:detector} line~\ref{ln:detector:enter} must be tagged as a
  \texttt{std::memory\_order\_seq\_cst} access to achieve
  requirement~\ref{rule:seq}, establishing a total order over these
  writes and subsuming release consistency.
  Line~\ref{ln:detector:waitingOn} must then be tagged
  \texttt{std::memory\_order\_acquire} to achieve
  requirement~\ref{rule:acq}.
  And finally, line~\ref{ln:detector:final} must be
  \texttt{std::memory\_order\_release} to satisfy~\ref{rule:rel}.
\end{itemize}

\subsection{Correctness}

Under the preceding consistency requirements, we can now prove
important theoretical guarantees of correctness for our deadlock
detector. Throughout, we consider an execution of a well-formed
program (\cref{def:wf}).

We first show that \cref{alg:detector} raises no false alarms.
\begin{theorem}
  If task $t$ fails the assertion in line~\ref{ln:detector:fail}
  during $\textsc{Get}(p)$, then a deadlock cycle exists, involving
  $t$ and $p$.
  \label{thm:precise}
\end{theorem}
\begin{proof}
  We have $t_0 = t$ and $p_0 = p$.
  If the execution had broken out of the while loop in
  line~\ref{ln:detector:breakt},~\ref{ln:detector:breakp},
  or~\ref{ln:detector:changed}, then the assertion would have
  succeeded.
  Therefore, it is the loop condition that fails.
  Upon reaching line~\ref{ln:detector:inc} in each iteration, we have
  found $p_i.\fldowner$ to be $t_{i+1}$ both before and after we found
  $t_{i+1}.\fldwaitingOn$ to be $p_{i+1}$. Therefore, we know 1) that
  at one time $t_{i+1}$ was the owner of $p_i$, and 2) that while
  $t_{i+1}$ still observed itself to own $p_i$, $t_{i+1}$ had invoked
  $\textsc{Get}(p_{i+1})$.
  This follows from memory consistency requirement~\ref{rule:acq}.
  At this point in the reasoning, we do not yet know if $t_{i+1}$
  still the owner of $p_i$ or if $t_{i+1}$ is still awaiting
  $p_{i+1}$.

  When the loop
  (lines~\ref{ln:detector:loop}--\ref{ln:detector:loopend}) terminates
  with $t_{i+1} = t_0$, since $t_0$ is the current task, we deduce
  that the final $t_{i+1}$, set by line~\ref{ln:detector:owner1}
  or~\ref{ln:detector:owner2}, is the current owner of $p_i$.
  For all $k$ modulo $i+1$, $t_k$ at one time concurrently observed
  itself to be the owner of $p_{k-1}$ and was in a call to
  $\textsc{Get}(p_k)$. This meets our definition of deadlock.
\end{proof}

The following series of lemmas builds to the theorem that
\cref{alg:detector} detects every deadlock.

\begin{definition}
  In a deadlock cycle comprising tasks $T$, a \emph{$t^*$ task} is a
  task in $T$ to which the line~\ref{ln:detector:enter} write by every
  task in $T$ is visible.
\end{definition}
\begin{lemma}
  Every deadlock cycle has a $t^*$ task.
  \label{lem:tstar}
\end{lemma}
\begin{proof}
  Corollary to memory consistency requirement~\ref{rule:seq}.
\end{proof}
A $t^*$ task, which need not be unique, should be thought of as the
(or a) last task to enter the deadlock.

\begin{lemma}
  If a program execution exhibits a deadlock cycle comprising tasks
  $T$ and promises $P$, when a $t^*$ task calls \textsc{Get} it
  constructs a sequence $\{ t_i \}_i$ that is a subset of $T$ and a
  sequence $\{ p_i \}_i$ that is a subset of $P$.
  \label{lem:inTP}
\end{lemma}
\begin{proof}
  We have $t_0 = t^* \in T$ and, by definition, $p_0 \in P$.
  If the loop immediately terminates, then $t_1 = t_0 \in T$, and we
  are done.
  Otherwise, the values of $t_{i+1}$ and $p_{i+1}$ inductively depend
  on $t_i$ and $p_i$.
  By definition of deadlock, one of the tasks in $T$, call it
  $o_{p_i}$, observes itself to be the owner of $p_i$. The most recent
  write to $p_i.\fldowner$ (recall all the writes are ordered by
  \cref{lem:writeorder}) occurred in program order before $o_{p_i}$'s
  line~\ref{ln:detector:enter} write.
  Therefore, memory consistency requirement~\ref{rule:seq} establishes
  that $t^*$ must read $t_{i+1} = o_{p_i} \in T$ in
  line~\ref{ln:detector:changed}.
  By definition of $t^*$ and by memory consistency
  requirement~\ref{rule:rel}, we see that
  line~\ref{ln:detector:waitingOn} observes $t_{i+1}$'s
  line~\ref{ln:detector:enter} write, not its
  line~\ref{ln:detector:final} write. Thus, $p_{i+1} \in P$ by
  definition of deadlock.
\end{proof}

\begin{lemma}
  If a program execution exhibits a deadlock cycle comprising tasks
  $T$, no $t^*$ task executes a diverging loop
  (lines~\ref{ln:detector:loop}--\ref{ln:detector:loopend}) in its
  call to \textsc{Get}.
  \label{lem:terminate}
\end{lemma}
\begin{proof}
  Suppose, during the call to \textsc{Get} by $t^*$, the loop does not
  terminate. Thus $t_i \ne t_0$ for any $i > 0$.
  But by \cref{lem:inTP}, the infinite sequence $\{ t_i \}_i$ is a
  subset of $T$.
  Therefore, $T$, in fact, exhibits a smaller cycle not involving
  $t_0$, violating the minimality condition in the definition of
  deadlock cycle.
\end{proof}

\begin{theorem}
  If a program execution exhibits a deadlock cycle comprising tasks
  $T$ and promises $P$, at least one task in $T$ fails the assertion
  in \cref{alg:detector} line~\ref{ln:detector:fail}.
  \label{thm:correct}
\end{theorem}
\begin{proof}
  Suppose for the sake of contradiction that a deadlock cycle arises and
  yet no assertion fails.
  So every task $t \in T$ enters the \textsc{Get} procedure and either
  blocks at line~\ref{ln:detector:return} on a promise in $P$ or
  diverges in an infinite loop.

  No task exits the loop by failing the loop condition,
  $t_{i+1} \ne t_0$, since this would directly fail the assertion in
  line~\ref{ln:detector:fail}.

  For each invocation of \textsc{Get} by a $t^*$ task, the loop cannot
  break in line~\ref{ln:detector:breakt} or
  line~\ref{ln:detector:breakp} because \cref{lem:inTP} implies no
  tasks or promises in the sequence are $\Null$.
  If the loop breaks in line~\ref{ln:detector:changed}, then $t^*$ has
  observed the owner of $p_i$ to change from one read to the
  next. This is impossible: both reads observe the current owner,
  $o_{p_i}$, by the same reasoning as in the proof of \cref{lem:inTP}.
  Finally, the loop cannot diverge for $t^*$, by \cref{lem:terminate}.
  Since there exists at least one $t^*$ task, by \cref{lem:tstar}, we
  have a contradiction.
\end{proof}

\begin{corollary}[to \cref{thm:precise,thm:correct}]
  \Cref{alg:detector} is precise and correct, guaranteeing the
  existence of a deadlock when an alarm is raised and raising an alarm
  upon every deadlock.
\end{corollary}

%% file: evaluation.tex
\section{Implementation and Evaluation}

We have implemented ownership semantics with omitted set and deadlock
detection in Java. We give a brief discussion of some of the practical
considerations in the design of this implementation. We then present
the results of a performance evaluation on a set of benchmark
programs.

\subsection{Objected-Oriented Promise Movement}

Introducing an explicit conception of ownership is minimally
disruptive. It is already the case that every promise is fulfilled by
at most one task, since two sets cause a runtime error. We only ask
that the programmer identify this task by leveraging the existing
structure of \kwasync directives.
However, for large, complex synchronization patterns that rely on many
promises, it can become tedious for a programmer to specify all the
relevant promises, one by one.

\lstChannel

In our Java implementation, an object-oriented approach can reduce the
burden of identifying which promises should be moved to new tasks.
In our Java implementation of these language features, classes
containing many promises may implement a \textsf{PromiseCollection}
interface so that moving a composite object to a new task is
equivalent to moving each of its constituent promises.
A channel class is shown in \cref{lst:channel}, illustrating that
complex and versatile primitives can be built on top of promises with
the aid of \textsf{PromiseCollection}.
This class behaves like a promise that can be used repeatedly, where
the $n$th \textsf{recv} operation obtains the value from the $n$th
\textsf{send} operation.
This behavior depends on dynamically allocated promises, and the
responsibility for the sending end of the channel is associated not to
the ownership of a single promise, but to the ownership of different
promises at different times. It is abstraction-breaking to ask the
channel user to manually specify which promise to move to a new task
in order to effectively move the sending end of the channel.
Instead, we give the impression that the channel object itself is
movable like a promise (line~\ref{ln:channel:b}), since it is a
\textsf{PromiseCollection}, and the implementation of \kwasync relies on
the \textsf{getPromises} method (line~\ref{ln:channel:a}) to
determine which promises should be moved.

\subsection{Exception Handling}

In an implementation of \cref{alg:owners}, some care must go into an
exception handling mechanism.
What code is capable of and responsible for recovering from the failed
assertion in line~\ref{ln:owners:async:E}?
And what happens if a task terminates early, with unfulfilled
promises, because of an exception?

Observe that line~\ref{ln:owners:async:E} occurs within an
asynchronous task after the user-supplied code for that task has
completed.
One solution is to add a parameter to \textsc{Async} so that the user
can supply a post-termination exception handler, which accepts the
list of unfulfilled promises, $t'.\fldowned$, as input.
Indeed, the fix for the AWS omitted set bug included such a mechanism
(not shown in \cref{lst:amazon})~\cite{AWSBugFixed}.
Alternatively, the runtime could automatically fulfill every
unfulfilled promise upon an assertion failure in
line~\ref{ln:owners:async:E}.
Some APIs, including in C++ and Java, provide an exceptional variant
of the completion mechanism for
promises~\cite{Cpp17,JavaCompletableFuture}.
In our implementation, we use this mechanism to propagate an exception
through the promises that were left unfulfilled.

Finally, observe that the correctness of \cref{alg:owners} only
depends on knowing when a task's $\fldowned$ list is empty. Therefore,
the $\fldowned$ list could be correctly replaced with a counter, which
would at least reduce the memory footprint of ownership tracking, if
not also the execution time of maintaining a list. However, doing so
would mean that an assertion failure in line~\ref{ln:owners:async:E}
could not indicate \emph{which} promises went unfulfilled. Therefore,
the implementation we evaluate uses an actual list.

\subsection{Benchmarks}

We evaluate the execution time and memory usage overheads introduced
by our promise deadlock detector on nine task-parallel programs. The
overheads are measured relative to the original, unverified baseline
versions.

\begin{enumerate}
\item Conway~\cite{ConwayBench} parallelizes a 2D cellular automaton
  by dividing the grid into chunks. We adapted the code from C to
  Java, using our \textsf{Channel} class (\cref{lst:channel}) in place
  of MPI primitives used by worker tasks to exchange chunk borders
  with their neighbors.

\item Heat~\cite{HeatBench} simulates diffusion on a one-dimensional
  surface, with 50 tasks operating on chunks of 40,000 cells for 5000
  iterations. Neighboring tasks again use \textsf{Channel} in place of
  MPI primitives.

\item QSort sorts 1M integers using a parallelized divide-and-conquer
  recursion; the partition phase is not parallelized. This is a
  standard technique for parallelizing Quicksort~\cite{QuicksortAlg}
  and has been previously implemented using the Habanero-Java
  Library~\cite{HJlib}. We implemented the finish construct, which
  awaits task termination using promises.

\item Randomized distributes 5000 promises over 2535 tasks spawned in
  a tree with branching factor of 3. Each task awaits a random promise
  with probability 0.8 before performing some work, fulfilling its own
  promises, and awaiting all its child tasks. We chose a random seed
  that does not construct a deadlock.

\item Sieve counts the primes below 100,000 with a pipeline of tasks,
  each filtering out the multiples of an earlier prime. A similar
  program is found in prior work~\cite{Ng16}.

\item SmithWaterman (adapted from HClib~\cite{hclib}; also used in
  prior work \cite{TJ,KJ}) aligns DNA sequences having 18,000--20,000
  bases. Each task operates on a $25 \times 25$ tile.

\item Strassen (such a program is found in the Cilk, BOTS, and KASTORS
  suites~\cite{Cilk,BOTS,Kastors}) multiplies sparse $128 \times 128$
  matrices containing around 8000 values. Divide-and-conquer recursion
  issues asynchronous addition and multiplication tasks, up to depth
  5.

\item StreamCluster (from PARSEC~\cite{Parsec}) computes a streaming
  $k$-means clustering of 102,400 points in 128 dimensions, using 8
  worker tasks at a time. We replaced the OpenMP barriers with
  promises in an all-to-all dependence pattern.

\item StreamCluster2 reduces synchronization in StreamCluster by
  replacing some of the all-to-all patterns with all-to-one when it is
  correct to do so. We also correct a data race in the original
  implementation.
\end{enumerate}

All benchmarks were run on a Linux machine with a 16-core AMD Opteron
processor under the OpenJDK 11 VM with a 1 GB memory limit.
A thread pool schedules asynchronous tasks by spawning a new thread
for a new task when all existing threads are in use. This execution
strategy is necessary in general for promises because there is no
\emph{a priori} bound on the number of tasks that can block
simultaneously.
We measured both execution time and, in a separate run, average memory
usage by sampling every 10 ms.
Each measurement is averaged over thirty runs within the same VM
instance, after five discarded warm-up runs; this is a standard
technique to mitigate the variability of JVM overheads, including JIT
compilation~\cite{Georges07}.

\tabResults

\begin{figure}
    \includegraphics[width=\columnwidth]{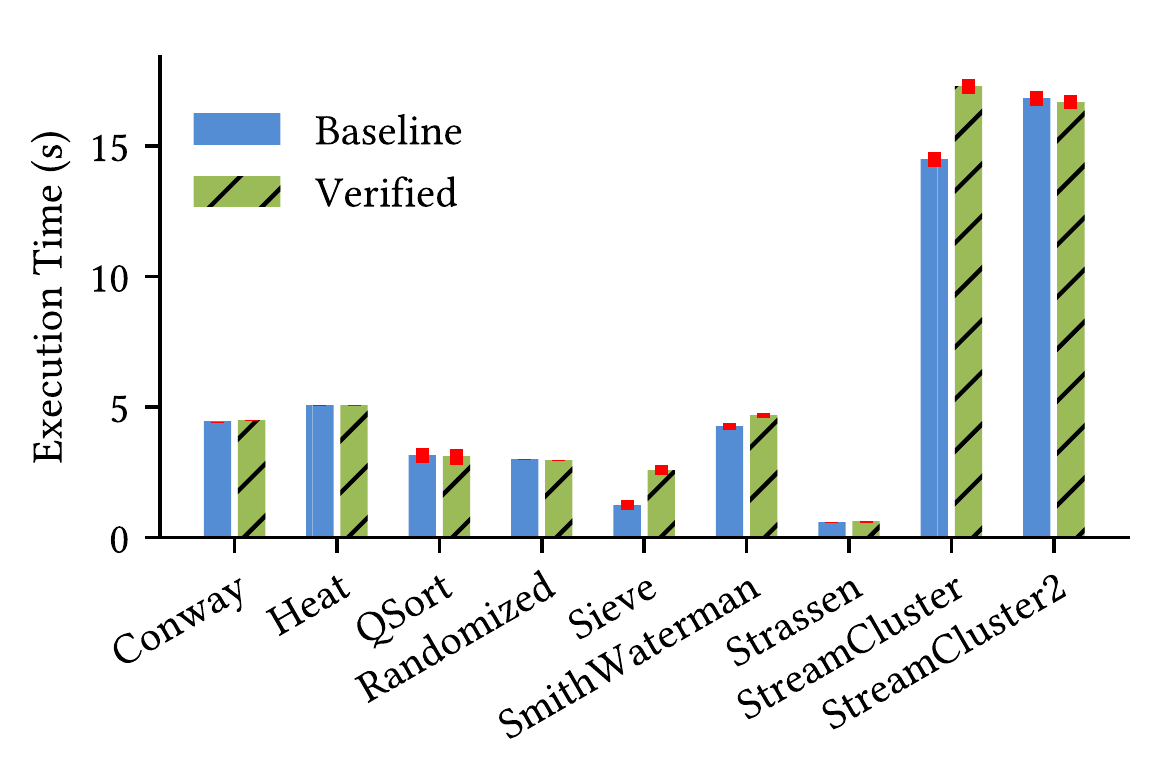}
    \caption{Execution times for each benchmark
      showing the mean with a 95\% confidence interval (red).}
    \label{fig:time}
    \Description{A plot of the baseline and verified execution times
      for each benchmark. The Sieve, SmithWaterman, and StreamCluster
      benchmarks have noticeable overheads.}
\end{figure}

\Cref{tab:results} gives the unverified baseline measurements for each
program and the overhead factors introduced by the verifiers.
The table also gives the geometric mean of overheads across all
benchmarks. There is an overall factor of \geomeanTime in execution
time and \geomeanMem in memory usage.
The total number of tasks in the program and the average rates of
promise get and set actions per millisecond (with respect to the
baseline execution time) are also reported.
\Cref{fig:time} represents the execution times of each benchmark,
showing the 95\% confidence interval.
The low overheads indicate that our deadlock detection algorithm does
not introduce serialization bottlenecks.

The overall execution time overheads are within 1.1$\times$ for each
of Conway, Heat, QSort, Randomized, SmithWaterman, Strassen, and
StreamCluster2. The same is true of the memory overheads for this
subset of benchmarks, excepting SmithWaterman. In many cases, the
verified run narrowly out-performs the baseline, which can be
attributed to perturbations in scheduling and garbage collection.

It is worth noting that the execution overhead for Sieve is in excess
of 2$\times$. Sieve has the single highest rate of get operations by
an order of magnitude (over 37,000, compared to SmithWaterman's
536). The Sieve program requires almost 9594 tasks to be live
simultaneously, each waiting on the next, with the potential to form
very long dependence chains for \cref{alg:detector} to traverse.

We can also remark on the 1.4$\times$ memory overhead in
SmithWaterman. Unlike Conway, Heat, Sieve, and both of the
StreamCluster benchmarks, in which most promises are allocated by the
same task that fulfills them, SmithWaterman (and Randomized) allocates
all promises in the root task and moves them later. In maintaining the
$\fldowned$ lists in \cref{alg:owners}, one can make trade-offs
between speed and space. Our implementation favors speed, so instead
of literally removing a promise $p$ from $t.\fldowned$ in
lines~\ref{ln:owners:async:Y} and~\ref{ln:owners:set:C}, we simply
rely on the fact that $p.\fldowner \ne t$ anymore to detect that $p$
should no longer be counted in line~\ref{ln:owners:async:E}.

For comparison with deadlock verification in other settings, the Armus
tool~\cite{Armus} can identify barrier deadlocks as soon as they
occur, with execution overheads of up to 1.5$\times$ on Java
benchmarks.
Our benchmark results represent an acceptable performance overhead
when one desires runtime-identifiable deadlocks and omitted sets with
attributable blame.

%% file: related.tex
\section{Related Work}

Task-parallel programming is prevalent in a variety of languages and
libraries.
Multilisp~\cite{Multilisp} is one of the earliest languages with
futures, a mechanism for parallel execution of functional code.
Fork-join parallelism is employed in Cilk~\cite{Cilk}, and the more
general async-finish with futures model was introduced in
X10~\cite{X10}.
Habanero-Java~\cite{HJ} modernized X10 as an extension to Java and,
later, as a Java library, HJlib~\cite{HJlib}; this language
incorporates additional synchronization primitives, such as the
phaser~\cite{Phaser} and the data-driven future~\cite{DDF}, which is a
promise-like mechanism.
Many other languages, libraries, and extensions include spawning and
synchronizing facilities, whether for threads or lightweight tasks,
including Chapel~\cite{Chapel}, Fortress~\cite{Fortress},
OpenMP~\cite{OpenMP}, Intel Threading Building Blocks~\cite{TBB},
Java~\cite{JCP}, C++17~\cite{Cpp17}, and Scala~\cite{ScalaFutures}.

The promise, as we define it, can be traced back to the I-structures
of the Id language~\cite{IStructures}, which are also susceptible to
deadlock.
Cells of data in an I-structure are uninitialized when allocated, may
be written to at most once, and support a read operation that blocks
until the data is available.

The classic definition of a deadlock is found in Isloor and
Marsland~\cite{Deadlock}, which is primarily concerned with concurrent
allocation of limited resources.
Solutions in this domain fall into the three categories of Coffman:
static prevention, run-time detection, and run-time
avoidance~\cite{Coffman71}.

We consider logical deadlocks, which are distinct from resource
deadlocks in that there is an unresolvable cyclic dependence among
computational results.
Solutions in the logical deadlock domain include techniques that
dynamically detect cycles~\cite{Luecke03, Krammer04, Krammer08,
  Hilbrich09, Vo11, Hilbrich12}, that raise alarms upon the formation
or possible formation of cycles~\cite{Agarwal06, Boudol09, Gerakios11,
  Armus, KJ, TJ}, that statically check for cycles through
analysis~\cite{Williams05, Naik09, Ng16} or through type
systems~\cite{Boyapati02,Vasconcelos09}, or that preclude cycles by
carefully limiting the blocking synchronization semantics available to
the programmer, either statically or dynamically~\cite{X10, Phaser,
  HJ, KJ, TJ}.
The present work includes a dynamic, precise cycle detection
algorithm, enabled only by the introduction of a structured ownership
semantics on the otherwise unrestricted promise primitive.

Futures are a special case of promises where each one is bound to a
task whose return value is automatically put into the promise.
Transitive Joins~\cite{TJ} and its predecessor, Known Joins~\cite{KJ},
are policies with runtime algorithms for deadlock detection on
futures. They are, in general, not applicable to promises. These two
techniques impose additional structure on the synchronization pattern
by limiting the set of futures that a given task may await at any
given time.

Recent work identifies the superior flexibility of promises over
futures with the problematic loss of a guarantee that they will be
fulfilled and develops a \emph{forward} construct as a
middle-ground~\cite{Forward}. Forwarding can be viewed in terms of
delegating promise ownership, but it is restricted in that 1) it moves
only a single promise into a new task, and 2) in particular, it moves
only the implicit promise that is used to retrieve a task's return
value. In terms of futures, forwarding amounts to re-binding a future
to new task.

Other synchronization constructs benefit from similar annotations to
the one we have proposed for promises. This includes event-driven
programming models where events have similar semantics to that of promises.
JavaScript, though a single-threaded language, still uses an
asynchronous task model to schedule callbacks on an event
loop~\cite{JavaScriptAsync}, and could benefit from our approach.
Likewise, our approach is directly applicable to multithreaded
execution models, such as Concurrent Collections~\cite{CnC} and the Open
Community Runtime~\cite{OCR}, that use event-driven execution as a fundamental
primitive.
As another example, the MPI blocking receive primitive must name the sending
task; from this information a waits-for graph for deadlock detection
can be directly constructed~\cite{Hilbrich09}.  In addition,
nonblocking communications in MPI
use {\tt MPI\_Request} objects in a manner similar to promises, and
the {\tt MPI\_Wait} operation akin to the get operation on promises.

Languages with barriers and phasers sometimes require the
participating tasks to \emph{register} with the
construct~\cite{Phaser}.
Notably, this kind of registration is absent from the Java API, which
is problematic for the Armus deadlock tool~\cite{Armus}. In that work,
registration annotations had to be added to the Java benchmarks in
order to apply the Armus methodology.

In this work, we considered programs which only use promises for
blocking synchronization, and we constrained ownership transfer to
occur only when a task is spawned.
Since a promise can have multiple readers or no readers at all, it is
not possible in principle to use one promise to synchronize the
ownership hand-off of a second promise between two existing tasks. We
cannot guarantee that the receiving task exists and is unique.
In future work, one could consider a slightly higher abstraction in
the form of a pair of promises acting like a rendezvous, which is a
primitive in languages like Ada and Concurrent
C~\cite{AdaConcurrentC}. Such a synchronization pattern could be
leveraged to hand off promise ownership since there would be a
guaranteed single receiving task.

The Rust language incorporates affine types in its move semantics to
ensure that certain objects have at most one extant reference at all
times~\cite{Rust}. The movement of promise ownership from one task to
another and the obligation to fulfill each promise exactly once may be
expressible at compile time through the use of a linear type system,
which restricts references to exactly one instance.